\documentclass[aps,pra,twocolumn,showpacs,10pt,superscriptaddress]{revtex4-1}
\usepackage{graphicx}
\usepackage{subfigure}
\usepackage{amsmath}
\usepackage{amsfonts}
\usepackage{amssymb}
\usepackage{mathrsfs}
\usepackage{bbm}
\usepackage{subfigure}
\usepackage{xcolor}

\begin{document}

\title{Manipulating the Majorana Qubit with the Landau-Zener-St\"{u}ckelberg Interference}
\author{Wen-Chao Huang}
\affiliation{
State Key Laboratory of Optoelectronic Materials and Technologies,
School of Physics and Engineering, Sun Yat-sen University, Guangzhou 510275, China}

\author{Qi-Feng Liang}
\affiliation{Department of Physics, Shaoxing University, Shaoxing 312000, China}

\author{Dao-Xin Yao}
\email{yaodaox@mail.sysu.edu.cn}
\affiliation{
State Key Laboratory of Optoelectronic Materials and Technologies,
School of Physics and Engineering, Sun Yat-sen University, Guangzhou 510275, China}

\author{Zhi Wang}
\email{wangzh356@mail.sysu.edu.cn}
\affiliation{
State Key Laboratory of Optoelectronic Materials and Technologies,
School of Physics and Engineering, Sun Yat-sen University, Guangzhou 510275, China}

\begin{abstract}
Constructing a universal operation scheme for Majorana qubits remains a central issue for the topological quantum computation. We study the Landau-Zener-St\"{u}ckelberg interference in a Majorana qubit and show that this interference can be used to achieve controllable operations.
The Majorana qubit consists of an rf SQUID with a topological nanowire Josephson junction which hosts Majorana bound states.
In the SQUID, a magnetic flux pulse can drive the quantum evolution of the Majorana qubit.
The qubit experiences two Landau-Zener transitions when the amplitude of the pulse is tuned around the superconducting flux quanta $2e/\hbar$.
The Landau-Zener-St\"{u}ckelberg interference between the two transitions rotates the Majorana qubit, with the angle controlled by the time scale of the pulse.
This rotation operation implements a high-speed single-qubit gate on the Majorana qubit, which is a necessary ingredient for the topological quantum computation.
\end{abstract}
\date{\today}
\maketitle


\section{Introduction}
Majorana bound states (MBSs) which reside in topological superconducting systems are drawing much attention both theoretically and experimentally\cite{kanermp,aliceareview,beenakkerreview,kitaev,law09,green,
kanefu,chliu,sarma1d,oppen1d,kouwenhoven,pergescience,flensberg,wang,beenakker13,lawnc,mengcheng}. The so-called topological superconductivity in this context is the spinless p-wave superconductivity, in which a Majorana number can be defined by the Pffafian of the Bogoliubov-de Gennes Hamiltonian\cite{kitaev}. MBSs are zero energy quasiparticles in these topological superconductors. They are localized at the ends of one dimensional systems\cite{kitaev}, or the core area of the superconducting vortices in two dimensional systems\cite{green}.
Electrons in natural systems always have spin, therefore, MBSs are predicted in artificial structures such as the interface of a three dimensional topological insulator and a superconductor\cite{kanefu,chliu}, the spin-orbit coupling nanowire in proximity to a superconductor\cite{sarma1d,oppen1d,kouwenhoven}, and the ferromagnetic atom chain on top of a superconductor\cite{pergescience}.
These systems have a common character that the spins of the electrons near the fermi surface are effectively eliminated by the spin-orbit coupling and the Zeeman energy. Then, the effective spinless superconductivity, {\it i.e.} the topological superconductivity, is achieved through the proximity effect.

MBSs in topological superconductors are interesting because of their non-Abelian exchange statistics\cite{ivanov,aliceanphy}. They provide a realistic platform for investigating a simplest example of non-Abelian particles.
Besides, MBSs are proposed to be useful in quantum computation\cite{kanermp}. They can construct Majorana qubits, which resist to local perturbations and can store quantum information for a long time.
Furthermore, the Majorana qubits can be rotated by the braiding of MBSs\cite{ivanov}. Importantly, these rotation operations are topologically protected. The effect of the operation is determined by the topology of the braiding, not the
detailed path. This topological quantum processing provides a possible scheme for the topological quantum computation. However, the braiding operations are not sufficient to realize universal quantum gates\cite{aliceareview}. They must be supplemented by non-topological operations on Majorana qubits, which have been proposed by using quantum dots\cite{flensberg,wang}, superconducting qubits\cite{jiang,bonderson}, or microwave cavities\cite{pekker}.

\begin{figure}[b]
\includegraphics[clip=true,width=1\columnwidth]{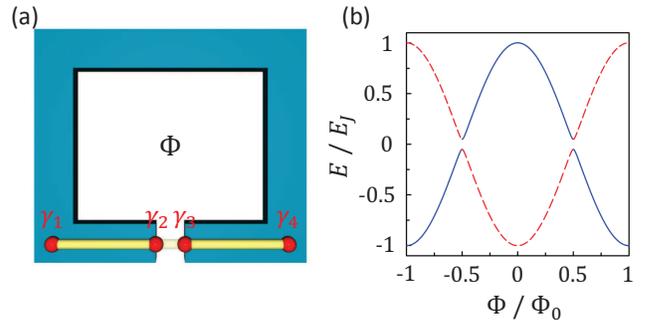}
\caption{(Color online). (a) Schematic setup of the Majorana qubit, which consists of an rf SQUID with a tunneling barrier junction on the topological nanowire. Four Majorana bound states are localized at the ends of the wire and the tunneling barrier. (b) The energy spectrum of the Majorana qubit as a function of magnetic flux $\Phi$, where the eigenstates are up-spin (solid line) and down-spin (dashed line) in the psudo-spin representation.}
\end{figure}

One promising approach for operating qubits is the Landau-Zener-St\"{u}ckelberg (LZS) interference\cite{nori}. The LZS interference
is a standard quantum phenomena in quantum mechanics. It occurs in quantum two-level systems in which the two levels exhibit avoided level crossings. At the two sides of the crossing point, the physical properties of the eigenstates are exchanged. Therefore, non-adiabatic Landau-Zener transitions between the two levels may happen when the system transverses the avoided crossing under a time varied control parameter\cite{landau,zener}. The interference between these transitions is called LZS interference\cite{stueckelberg}.
The LZS interference is very useful for coherent quantum operation, since it is robust to certain noises and possible to implement high fidelity quantum gates\cite{nori,GGC}. In Ref. [27], the LZS interference has been experimentally
achieved for realistic quantum control on quantum dot qubits.

In this work, we propose to use the LZS interference to implement quantum operations on Majorana qubits. For this purpose, we adopt the system sketched in Fig. 1, which consists of an rf SQUID with a topological nanowire Josephson junction. The nanowire hosts MBSs near the tunneling barrier and the ends of the wire.
These MBSs form a Majorana qubit, which is correlated with the superconducting phase difference across the junction. Because of this correlation, the Majorana qubit is described by the Josephson energy of the junction, which is a $2\times2$ Hamiltonian with avoided level crossings.
If a magnetic flux pulse is applied, the Josephson Hamiltonian will evolve and
transverse through these crossings. We consider a triangular pulse, for which the maximum flux value is tuned around the superconducting flux quanta $\Phi_0=2e/\hbar$. Within one pulse, the Josephson Hamiltonian transverses through the same crossing point twice, and then come back to the initial Hamiltonian. In contrast, the Majorana qubit state does not come back to the initial state after the pulse.
Landau-Zener transitions happen at the avoided crossings, and the resulted LZS interference rotates the Majorana qubit, where the rotating angle is controlled by changing the time length of the pulse. Our work provides a one-qubit gate on the Majorana qubit, which is important for realizing topological quantum computation.

We organize this work as follows. The model for the system is presented in section II, then we study the LZS interference analytically under this model in section III. We present a numerical simulation in section IV. Finally we give discussions and a conclusion in section V.

\section{Model}
The system illustrated in Fig. 1a is constructed by a nanowire Josephson junction which hosts four MBSs. Two of them ($\gamma_1$, $\gamma_4$) are localized at the two ends of the wire, and the other twos ($\gamma_2$, $\gamma_3$) at the two sides of the tunneling barrier. The Majorana qubit built by the two MBSs near the tunneling barrier is directly described by the Josephson energy of the junction,
which is different from conventional junctions in two aspects. First, it is actually a matrix for a Majorana junction\cite{kitaev}, rather than a pure number for conventional junction. The basis states of the matrix are the Majorana qubit states. In this sense, it is more accurate to be named as Josephson Hamiltonian. Second, the Josephson Hamiltonian has the $4\pi$ period components in its diagonal elements\cite{kitaev,aliceareview}, different from the $2\pi$ period Josephson energy in conventional junctions. With these insights, the Josephson Hamiltonian for this Majorana junction has been given as\cite{aliceareview},
\begin{equation}
H = iE_J\gamma_2\gamma_3\cos({\pi \Phi }/{\Phi_0}) + i\delta_L\gamma_1\gamma_2 + i\delta_R\gamma_3\gamma_4,
\end{equation}
where $\gamma_{1,2,3,4}$ are the four MBSs, $E_J$ is the Josephson energy due to the coupling of the two MBSs around the tunneling barrier; $\delta_{L,R}$ represent the coupling between the distant MBSs at the left and the right side of the barrier, respectively; $\Phi$ is the applied magnetic flux through the SQUID, which controls the phase difference across the junction in the vanishing inductance regime under current consideration.

We define two fermionic operators $f^\dagger_1 = (\gamma_2 + i\gamma_3)/2$ and $f^\dagger_2 = (\gamma_4 + i\gamma_1)/2$ to construct a fermionic representation for the MBSs. Then the Hamiltonian is transformed to,
\begin{equation}
\begin{aligned}
H=& E_J(1 - 2f^\dagger_1f_1)\cos{({\pi \Phi}/{\Phi_0})} + \delta_+(f^\dagger_2f_1 + f^\dagger_1f_2)\\
&+ \delta_-(f^\dagger_2f^\dagger_1 + f_1f_2).
\end{aligned}
\end{equation}
where $\delta_{\pm} = \delta_L \pm \delta_R$.
We find that the Josepshon Hamiltonian is depending on the occupation states of the two fermionic operators, which is nothing but the parity states built by the four MBSs\cite{pekker}. We are studying the quantum coherent evolution of the Majorana states, which occurs at a low temperature where the superconducting energy gap becomes large enough. Thus we can ignore the quasiparticle poisoning from high energy quasiparticles which strongly suppressed by the superconducting gap at low temperature.
In this case,
the total fermionic parity is conserves in this system. The Hilbert space of the MBSs can be divided into two disconnected subspace with even and odd total fermionic parities. Without losing generality, we choose the even total parity subspace. In this subspace, we can define psudo-spin states $\mid \uparrow \rangle = |0\rangle$ and $\mid \downarrow \rangle = f^\dagger_1 f^\dagger_2|0 \rangle$, with $\mid 0\rangle$ the vacuum states for the two fermionic operators.
These two psudo-spin states represent the two eigenstates of the Majorana qubit. Therefore, a spin representation for the Majorana qubit is established. In this representation, the Hamiltonian of the Majorana qubit is rewritten as,
\begin{equation}\label{spin}
H = E_J\cos\left({\pi\Phi}/{\Phi_0}\right)\sigma_z + \delta \sigma_x,
\end{equation}
where $\sigma_{x,z}$ are Pauli matrices acting on the pusdo-spin states, and $\delta=\delta_-$. The two eigenvalues of $H$ are $E_\pm = \pm \sqrt{E^2_J \cos^2(\frac{\pi\Phi}{\Phi_0}) + \delta^2}$ , which are depicted in Fig. 1b as a function of magnetic flux $\Phi$.
This Hamiltonian is a typical two-level qubit system with avoided level crossings. Away from the crossings, the coupling between spin-up and spin-down state is small compared to the energy difference. Then these two spin states are the eigenstates of the qubit. The qubit dynamics is adiabatic. However, the energy difference between the two spin states vanishes at the crossing points, then the spin coupling $\delta$ dominates the dynamics of the qubit states. When the system traverses through the crossings under a time varying magnetic flux, interesting physics occurs: non-adiabatic Landau-Zener transitions may happen, and the qubit state will experience a rotation on the Bloch sphere representation\cite{nori}.
This Landau-Zener transition has been extensively studied in the literature\cite{nori}, with realistic examples achieved in atomic systems, quantum dot systems, and superconducting systems.

\section{Landau-Zener-St\"{u}ckelberg Interference}
The Landau-Zener transition can significantly modulate the quantum state of the Majorana qubit. However, the system will not restore to its original Hamiltonian after one Landau-Zener transition. The Hamiltonian
will be added with magnetic flux, which brings an obstacle for the realistic quantum control on the Majorana qubit.
One solution to this obstacle is to consider a magnetic flux pulse. During the pulse, the flux first increases and the system transverses through an avoided crossing point, then the flux decreases and the system transverses back through the same crossing point. Finally, the flux vanishes after the pulse and the system will restore to the original Hamiltonian. The Majorana qubit transverses through the same crossing point twice within one pulse. Two Landau-Zener transition happens and the LZS interference will rotate the Majorana qubit.

A flux pulse with a triangular shape is a natural choice for this purpose. It is one of the simplest pulse shapes, and provides a constant increasing and decreasing speed for the magnetic flux. This linear dependence simplifies the analytic solutions. We consider an triangular pulse with a function of,
\begin{equation}
\Phi(t) = \Phi_0 \times
\begin{cases}
\omega_1 t, & 0 < t <\frac{A}{\omega_1}\\
A - \omega_2 (t - \frac{A}{\omega_1}), & \frac{A}{\omega_1} < t < \frac{A}{\omega_1} + \frac{A}{\omega_2}\\
0, & others
\end{cases}
\end{equation}
where the amplitude $A$ and the velocity $\omega_{1,2}$ are positive. As shown in Fig. 2a, this pulse starts from time zero, increases with a constant speed $\omega_1$, reaches a maximum value of $A\Phi_0$, then decrease with a constant speed of $\omega_2$, finally vanishes at time ${A}/{\omega_1} + {A}/{\omega_2}$. This triangular pulse is asymmetric for general case. It becomes symmetric under the special condition of $\omega_1 = \omega_2$. In the following calculations, we find that the asymmetric pulse is more suitable for our purpose to control the Majorana qubit.

\begin{figure}[t]
\includegraphics[clip=true,width=1\columnwidth]{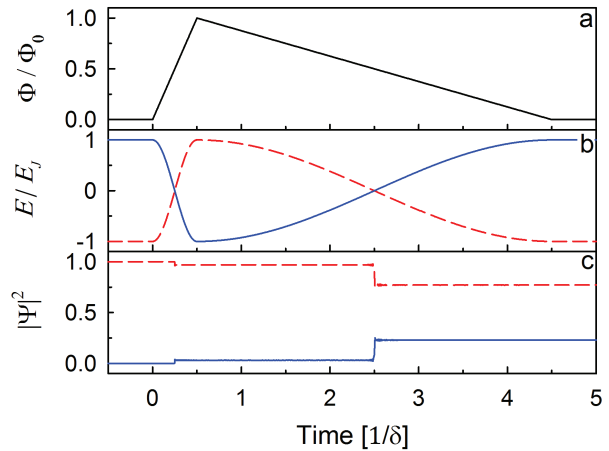}
\caption{(Color online). (a) The asymmetric triangular flux pulse as a function of time. (b) The time evolution of the two diagonal terms of the Hamiltonian Eq. (\ref{spin}). (c) The time evolution of the Majorana qubit sate represented by the two components $|\psi_1(t)|^2$ (solid line) and $|\psi_2(t)|^2$ (dashed line). The parameters are $A = 1$, $\delta/E_J = 0.001$, $\omega_1/\delta = 0.2$ and $\omega_2/\delta = 0.025$.}
\end{figure}

Landau-Zener transitions appear in this system only when the pulse amplitude $A$ is greater than $1/2$. For simplicity, we consider a pulse amplitude of $1/2 < A < 3/2$. Then only one avoided crossing point is involved during the pulse. In the following analysis, we denote $t_1 = {1}/{(2\omega_1)}$ and $t_2 = t_1 + (A - 1/2)/\omega_2$ as the crossing time when the system transverses across the avoided crossing $\Phi = \Phi_0/2$.
Near $t_1$ and $t_2$, the diagonal part of the Hamiltonian is small and can be linearized as\cite{nori},
\begin{equation}
\epsilon(t) = E_J \cos(\pi \Phi(t) /\Phi_0) \approx \pm v_{1,2} (t - t_{1,2}),
\end{equation}
where
\begin{equation}
v_{1,2} = E_J \left[\frac{d}{dt}\cos({\pi\Phi(t)}/{\Phi_0})\right]_{t = t_{1,2}} = \pi E_J \omega_{1,2}.
\end{equation}
Then the effective Hamiltonian is linearized near the crossing times,
\begin{equation}
H_{1,2}(t) = v_{1,2} (t - t_{1,2}) \sigma_z + \delta \sigma_x.
\end{equation}
These linearized Hamiltonian $H_{1,2}$ are typical examples of Landau-Zener problem. The Landau-Zener transition probability for a single crossing is given by\cite{landau,zener}
\begin{equation}
P_{1,2} = \exp(-2\pi\beta_{1,2}),
\end{equation}
where $\beta_{1,2} = \delta^2/2v_{1,2}$. If the system starts from the lower level initially, then the Landau-Zener transition probability $P_{1,2}$ describes the probability of upper level occupation after one crossing.
As shown in Ref. [28], these Landau-Zenner transitions can be described approximately by an unitary evolution matrix $\hat{N}_{1,2}$\cite{nori,nori2},
\begin{equation}
\begin{aligned}
&\hat{N}_{1,2} = \left(\begin{array}{cc}
\sqrt{P_{1,2}} & \sqrt{1 - P_{1,2}}e^{i\tilde{\varphi}_{1,2}}\\
-\sqrt{1 - P_{1,2}}e^{-i\tilde{\varphi}_{1,2}} & \sqrt{P_{1,2}}
\end{array}\right),
\end{aligned}
\end{equation}
where $\tilde{\varphi}_{1,2} = -\frac{\pi}{2} + \beta_{1,2}(\ln\beta_{1,2} - 1) + arg \Gamma(1 - i\beta_{1,2})$, $\Gamma$ is the gamma function and the phase $\tilde\varphi_{1,2}$ is monotonous function changes from 0 in the adiabatic limit ($P_{1,2} \rightarrow 0$) to $\pi/4$ in the diabatic limit ($P_{1,2} \rightarrow 1$)\cite{nori}.
These two evolution matrices connect the wave function of the system before and after the Landau-Zener transitions,
which reads $\Psi(t_{1,2}+0) = \hat{N}_{1,2}\Psi(t_{1,2}-0)$ with $\Psi= \psi_1 \mid \uparrow \rangle + \psi_2 \mid \downarrow \rangle$.

Now we consider the LZS interference between the two Landau-Zener transitions, and obtain the finial Majorana qubit state. The system evolution between these two Landau-Zener transitions is approximated by an adiabatic evolution matrix\cite{nori,nori2},
\begin{equation}
\hat{U}(t',t) = \left(\begin{array}{cc}
e^{-i\zeta(t',t)} & 0 \\
0 & e^{i\zeta(t',t)}
\end{array}\right),
\end{equation}
where $\zeta(t',t) = \frac{E_J}{\hbar}\int^{t'}_t \cos{\frac{\pi \Phi(\tau)}{\Phi_0}} d\tau$ are adiabatic evolution phase.
The quantum state rotation induced by the LZS interference is written as\cite{GGC},
\begin{equation}\label{final_Psi}
\Psi (t) \approx \hat{U}(t,t_2)\hat{N}_2\hat{U}(t_2,t_1)\hat{N}_1\hat{U}(t_1,t_0)\Psi_0,
\end{equation}
where $\Psi_0 = (0,1)^T$ is the initial state for the Majorana qubit, and $\Psi(t)= [\psi_1(t),\psi_2(t)]^T$ is the final state.
We take $\hat{U}(t_1,t_0)$ as identity matrix with appropriate choice of the phase for the basis states. Then the final qubit state reads,
\begin{eqnarray}
\psi_1(t) && = [\sqrt{P_1(1 - P_2)} e^{i\zeta(t_2,t_1) + i\tilde{\varphi_2}} \\\nonumber
&&+ \sqrt{(1 - P_1)P_2} e^{-i\zeta(t_2,t_1) + i\tilde{\varphi_1}}]e^{-i\zeta(t,t_2)}.
\end{eqnarray}
We immediately find that the Majorana qubit state has been rotated. The phase of $\psi_1(t)$ experiences the Larmor precession, therefore will rotate after the pulse horizontally in the Bloch sphere. In principle, any horizontal rotation angle for the Majorana qubit can be achieved with appropriate precession time\cite{GGC}. Here, we concentrate on the longitudinal rotation of the Majorana qubit, which is given by the amplitude of the final wave function,
\begin{eqnarray}\label{upper}
|\psi_1|^2 &&= (P_1 + P_2 - 2P_1 P_2) \\\nonumber
&&+ 2\sqrt{P_1 P_2 (1 - P_1)(1 - P_2)} \cos \chi,
\end{eqnarray}
where $\chi = 2\zeta(t_2,t_1) - \tilde{\varphi}_1 + \tilde{\varphi}_2$ comes from the adiabatic phase evolution, which is
sensitive to the parameters of the pulse\cite{nori}. The first term in Eq. (\ref{upper}) provides the rotation angle of the Majorana qubits from $(0,1)$ to $[\psi_1(t), \psi_2(t)]$. The rotation angle is fully determined by the probability of the two Landau-Zener transitions $P_1$ and $P_2$, therefore is controlled by the flux varying speed $\omega_1$ and $\omega_2$.
The second term can be treated as
the uncertainty for the rotation operation, since the phase factors are uncontrollable.
We could reduce this uncertainty by modulating $P_1$ or $P_2$, making one of them approaches $0$ or $1$. This regime can be achieved only for the asymmetric triangular pulse, since the symmetric triangular pulse leads to $P_1 = P_2$. Then the uncertainty term will always be significant.

\begin{figure}[tb]
\includegraphics[clip=true,width=1\columnwidth]{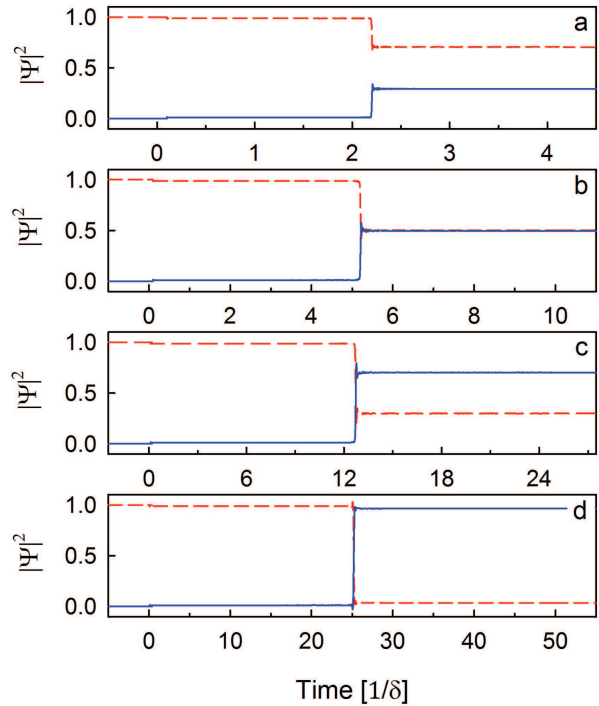}
\caption{(Color online) The time evolution of the Majorana qubit sate represented by the two components $|\psi_1(t)|^2$ (solid line) and $|\psi_2(t)|^2$ (dashed line), with (a) $\omega_2/\delta = 0.025$, (b) $\omega_2/\delta = 0.01$, (c) $\omega_2/\delta = 0.004$, and (d) $\omega_2/\delta = 0.002$. Other parameters are $A = 1$, $\delta/E_J = 0.001$, and $\omega_1/\delta = 0.5$}
\end{figure}

\section{Numerical simulation}
Now we study the system described by Eq. (3) numerically to obtain quantitative results.
First we setup the equations for numerical simulations. We expand the Hilbert space of the Majorana qubit system with the psudo-spin states $|\Psi (t) \rangle = \psi_1(t) \mid\uparrow \rangle + \psi_2 (t) \mid \downarrow \rangle$. Then the Schr\"{o}dinger equation for the Majorana qubit is written explicitly as,
\begin{equation}
i\frac{d}{dt}  \left(\begin{array}{c}
\psi_1\\
\psi_2
\end{array}\right) = \left(\begin{array}{cc}
E_J\cos\frac{\pi\Phi(t)}{\Phi_0} & \delta\\
 \delta& -E_J\cos\frac{\pi\Phi(t)}{\Phi_0}
\end{array}\right)
\left(\begin{array}{c}
\psi_1\\
\psi_2
\end{array}\right).
\end{equation}
Numerical simulations for this Schr\"{o}dinger equation are performed with standard finite difference method, in which the evolution operation is linearized within each small segment of time. We consider the asymmetric triangular magnetic flux pulse described in Eq. (4). The signal starts from $t=0$ and ends at $t=A/\omega_1 + A/\omega_2$, as shown in Fig. 2a. This pulse increases the flux through the SQUID from zero to $A\Phi_0$ rapidly, then reduces back to zero slowly. During this process, the two diagonal terms of the Hamiltonian, which is shown in Fig. 2b, variate and cross with each other twice.
Around these two special crossing points, the diagonal part vanishes, thus the off-diagonal part dominate the system and the qubit experiences Landau-Zener transitions. We calculate the wave function evolution of the Majorana qubit starting from $\Psi=(0,1)^T$, and show the results in Fig. 2c. The qubit is staying at the initial state before the pulse. After $t=0$, the pulse is applied and the Hamiltonian begins evolution. We find that the system stays at the initial state away from the crossing points, because
the energy difference between the diagonal part of the Hamiltonian is big enough to induce adiabatic dynamics. When the flux increases and the system approaches the avoided crossing, the evolution becomes non-adiabatic, and the qubit
will have a probability of psudo-spin rotation. This is the Landau-Zener transition as expected in theoretical analysis. The increasing speed of the flux is very quick for this pulse. Therefore, the evolution is in the diabatic limit. The psudo-spin rotation is small.
When the flux decreases, the second Landau-Zener transition happens. This time, the decreasing speed of the flux is moderate. The LZS interference between these two transitions then rotates the qubits, which is clearly demonstrated with the two components of the wave function in Fig. 2c.

This qubit rotation is controllable by modulating the time scale of the magnetic pulse. More specifically, it is controlled by the flux decreasing speed of the pulse, since we increase the flux fast enough to achieve the extreme diabatic behavior. In this scenario, the error term due to the phase accumulation in the adiabatic region is reduced.
We show the results with different rotation angles for the Majorana qubit in Fig. 3. The rotation angle is determined by the flux decreasing speed of the pulse $\omega_2$.
We find that the rotation can be as large as an inversion from $(0,1)$ state to $(1,0)$ state.
With these rotation operations, we achieve a one-qubit gate for the Majorana qubit simply by applying a magnetic flux pulse.
This one-qubit gate is a good supplement to the braiding operations, and should be important for realizing universal quantum gates in topological quantum computation.

\section{Discussions and Conclusion}
In this work, we exploit the asymmetric triangular pulse to implement the one-qubit control. The triangular pulse has the advantage of being simple in theory, yet is experimentally easy to achieve. However, a symmetric pulse causes a large error as shown in Eq. (\ref{upper}), and the operation on the qubit will be extremely sensitive to the parameters of the pulse. Therefore,
it is difficult to achieve a realistic control on the Majorana qubit. In contrary, the asymmetric pulse can induce well controlled one-qubit gate, in which the qubit rotation angle is fully determined by the decreasing speed of the pulse. Therefore, it is more applicable for realistic quantum gates.

Finally, we discuss the orders of the physical quantities in our work. The topological superconductivity is achieved through the proximity effect, with an approximate critical temperature of  $T_c \sim $ 10 K. The Josephson energy $E_J$ should be much smaller than the superconducting gap, with a typically value of $E_J \sim $ 10 GHz\cite{pekker}. The coupling energy between distant Majorana bound states are exponentially small for long wire, which can be reasonably taken as $\delta \sim $ 10 MHz. Then the pulse length for operation Majorana qubits should be in the range of $\omega_{1,2} \sim$ (1-10) MHz. With these parameters, we estimate the time scale of the operations to be around (0.1-1) ms, which establishes a high-speed quantum gate on Majorana qubits.

In summary, we analyze the Landau-Zener-St\"{u}ckelberg interference of a Majorana qubit in a topological rf SQUID. An asymmetric triangular magnetic flux pulse is applied in the SQUID to drive the system. In one pulse period, the system transverses through the same avoided crossing point twice with different speed, and two Landau-Zener transitions happen. The Landau-Zener-St\"{u}ckelberg interference between these two transitions induces a rotation on the Majorana qubit. Importantly, the rotation angle can be controlled by the time scale of the pulse. Therefore, the Landau-Zener-St\"{u}ckelberg interference can achieve a one-qubit gate for the Majorana qubit. This quantum gate might be useful for topological quantum computation.

\section*{Acknowledgment}
This work was supported by NSFC-11304400, NSFC-61471401 SRFDP-20130171120015, 985 Project of Sun Yat-Sen University, and State Key Laboratory of Optoelectronic Materials and Technologies.
D.X.Y. is supported by NSFC-11074310, NSFC-11275279, SRFDP-20110171110026, NBRPC-2012CB821400, and Fundamental Research Funds for the Central Universities of China.

\end{document}